\documentclass[12pt]{article}
\usepackage{graphicx}
\usepackage{amssymb,amsmath,amsfonts,palatino,amsthm}
\usepackage{amssymb}
\usepackage{epstopdf}
\newcommand{\beq}{\begin{equation}}
\newcommand{\eeq}{\end{equation}}

\newcommand{\f}{\begin{equation}}
\newcommand{\ff}{\end{equation}}

\newcommand\be{\begin{equation}}
\newcommand\ba{\begin{eqnarray}}
\newcommand\ee{\end{equation}}
\newcommand\ea{\end{eqnarray}}

\begin{document}

\title{Gravitational origin of the weak interaction's chirality \\}
\author{Stephon Alexander\thanks{stephon.alexander@dartmouth.edu} , Antonino Marcian\`o\thanks{antonino.marciano@dartmouth.edu}
\\
Department of Physics \& Astronomy, HB 6127\\
Wilder Lab, Dartmouth College, Hanover, NH 03755, USA
\\
\\
Lee Smolin\thanks{lsmolin@perimeterinstitute.ca} 
\\
Perimeter Institute for Theoretical Physics,\\
31 Caroline Street North, Waterloo, Ontario N2J 2Y5, Canada}
\date{\today}
\maketitle

\begin{abstract}
\noindent We present a new unification of the electro-weak and gravitational interactions based on the joining the weak $SU(2)$ gauge fields with the left handed part of the space-time connection, into a single gauge field valued in the complexification of the local Lorentz group.  Hence, the weak interactions emerge as the right handed chiral half of the space-time connection, which explains the chirality of the weak interaction. This is possible, because, as shown by Plebanski, Ashtekar, and others, the other chiral half of the space-time connection is enough to code the dynamics of the gravitational degrees of freedom.

This unification is achieved within an extension of the Plebanski action previously proposed by one of us.  The theory has two phases.  A parity symmetric phase yields, as shown by Speziale, a bi-metric theory with eight degrees of freedom: the massless graviton, a massive spin two field and a scalar ghost.  Because of the latter this phase is unstable.  Parity is broken in a stable phase where the eight degrees of freedom arrange themselves as the massless graviton coupled to an $SU(2)$ triplet of chirally coupled Yang-Mills fields.  It is also shown that under this breaking a Dirac fermion expresses itself as a chiral neutrino paired with a scalar field with the quantum numbers of the Higgs.  

\end{abstract}

\tableofcontents

\section{Introduction}

The ambition of unifying gravity with the other interactions faces three big obstacles:

\begin{enumerate}

\item{}Gravity is described by a dynamical metric while the other interactions are described by connection fields.  Consequently the Einstein action
is linear in curvature while the Yang-Mills action is quadratic in gauge field strength.

\item{}The standard model can be quantized perturbatively, because its action is a polynomial of dimension four terms, while the Einstein-Hilbert action, being non-polynomial, is challenging to quantize.  

\item{}The standard model of particle physics is chiral,
while gravity, at least at the classical level, is not.  Any unification must explain why parity is broken only for the weak interactions.

\end{enumerate}

The first  two challenges are addressed by the Ashtekar-Plebanski formulations of general relativity in which gravity is described by a gauge field \cite{Plebanski:1977zz, Ashtekar:1986yd}, while the metric is emergent \cite{Capovilla:1991kx, Capovilla:1991qb}.  These connection formulations of gravity are drastically simpler than Einstein's original metric formulation, as the action and hamiltonian formulations are based on cubic polynomials in the basic fields, which is a much better situation for quantization than Einstein's non-polynomial formulation.  Indeed these theories are as simple as non-linear theories can be, with purely quadratic field equations.

Remarkably,  these connection formulations of gravity address the issue of chirality as well.
There are a range of these Ashtekar-Plebanski formulations, which differ in the value of a complex parameter-the Immirzi parameter, $\gamma$.
When $\gamma$ takes complex values the action for gravity is chiral.  At the classical level this chirality is hidden in the gravitational sector and
affects only four fermion interactions that
arise from their couplings to the torsion of the connection.  But the chirality emerges in the quantum theory \cite{Perez:2005pm, Freidel:2005sn} where it can cause parity breaking
in the production of tensor modes in inflation \cite{Joao}. This could be detected as correlations of $B$ mode polarization with temperature fluctuations \cite{Contaldi:2008yz}.

The gravitational action is maximally chiral when $\gamma$ is purely imaginary in the sense that the gravitational action is then just a function
of the left handed part of the space-time connection.  Hence the connection and curvature that arise in the gravitational action and field equations are  valued only in $SU(2)_L$.  Any dependence on $SU(2)_R$ drops out.   The parity invariance of the classical Einstein equations arise from 
the fact that the complex conjugate of the left handed part can be inserted into the expressions of the equations of motion without changing their
on shell solutions.

The fact that the Einstein equations can be generated by an action which involves only the chiral $SU(2)_L$ half of the space-time connection \cite{Jacobson:1987yw} opens the door to an idea about unification: {\it perhaps the initial action for gravity is parity symmetric, but there is a phase in which parity is broken so that one chiral half, $SU(2)_L$, of the space-time connection codes the gravitational interactions, while the other chiral half, $SU(2)_R$, emerges with the dynamics of a Yang-Mills field propagating on a space-time described by the left half of the connection.}  There might then also be a phase in which parity is restored so that  both chiral halves carry gravitational dynamics.  

Note that this idea requires doubling the degrees of freedom initially, for in general relativity, applied to real, Lorentzian metrics, the left and right halves of the space-time connection are complex conjugates of each other.  To free them up these {\it reality conditions} have to be lifted, and replaced by alternative reality conditions which allow the  left and right halves of the connection to be independent of each other, but in a way which still realizes the reality of the metric.  We have  a proposal for these alternative reality conditions, but before introducing them we have to introduce the degrees of freedom that make it possible to realize our scenario.

Our starting point is a gauge theory of the complexified Lorentz group, $SL(2,\mathbb{C})_\mathbb{C}$ on a four dimensional manifold $\cal M$. The space-time connection $A^{ab}=-A^{ba}$ is a one form, valued in $\mathfrak{sl}(2,\mathbb{C})_\mathbb{C}$ the Lie algebra of $SL(2,\mathbb{C})_\mathbb{C}$.  That Lie algebra is represented by complex antisymmetric, $4 \times 4$ matrices, $M^{ab}=-M^{ba}$, where $a,b,c=0,1,2,3$ are internal Lorentz indices.   Because we want the metric to be emergent, we do not include it as a fundamental degree of freedom.  Instead, we write dynamics of $A^{ab}$ making use of two auxiliary fields, a two form $B^{ab}$, also valued in the Lie algebra of $SL(2,\mathbb{C})$ and a scalar field which provides a map $\Psi : \mathfrak{sl}(2,\mathbb{C})_C \rightarrow \mathfrak{sl}(2,\mathbb{C})_C$, which is written as
$\Psi_{abcd}$ with the following symmetries and constraints,
\f
\Psi_{abcd} =\Psi_{cdab} = - \Psi_{bacd}, \ \ \ \ \ \varepsilon^{abcd} \Psi_{abcd} =0
\ff
To specify the dynamics we choose the most general parity symmetric\footnote{With parity transformations applied simultaneously to space-time and  internal Lorentz indices.} polynomial of dimension four and less:
\begin{eqnarray}
S&=& \int \frac{1}{8 \pi G} \left \{ \varepsilon_{abcd}\, B^{ab} \wedge F^{cd} - \frac{1}{2} \Psi_{abcd} \, B^{ab} \wedge B^{cd} \right \} \nonumber \\
&&
+ \left( \frac{\Lambda}{16\pi G} -  \frac{g^2}{2}   \Psi_{abcd}^2 \right)\varepsilon_{efgh}\, B^{ef}\wedge B^{gh} 
+ \frac{\alpha}{2} \varepsilon_{abcd}\, F^{ab} \wedge F^{cd}\,,
 \label{action0}
\end{eqnarray}
where $F^{ab}$ is a two form which is the field strength of $A^{ab}$,  $G$ is Newton's constant and $\Lambda$ is the cosmological constant.
$A^{ab}$ then naturally has dimensions of inverse length, $B^{ab}$ is dimensionless and $\Psi_{abcd}$ has dimensions of inverse length squared. $\Psi_{abcd}^2=\Psi_{abcd}\, \Psi^{abcd}$, $\varepsilon^{abcd}$ is the Levi-Civita symbol and $g$ is a new dimensionless coupling constant.  Note that there is no Immirzi parameter as we restrict the action to parity even terms.  

The last term is a topological invariant.  Apart from that there is only a single term with a derivative in it, which is the first term.  

This action has been studied in several forms.  Without the terms in $\Psi$, it describes $BF$ theory, a topological theory \cite{BF}.  With $g^2=0$ it is a form of the Plebanski action for general relativity \cite{Plebanski:1977zz}.  The full action gives an extended dynamics for the gravitational field as discussed in \cite{Smolin}.  It has been studied by Alexandrov and Krasnov \cite{sergei-kirill} and Speziale \cite{Speziale:2010cf} and is known to have eight degrees of freedom. Alexandrov and Krasnov and Speziale studied the symmetric phase and found a bi-metric theory with a massless graviton, a massive spin two field and a scalar ghost $(8=2+5+1)$. The presence of the scalar ghost might have been suspected from results of Berezhiani, Comelli, Nesti and Pilo in \cite{Berezhiani:2007zf}, which show it bedevils a large class of bi-metric theories. 

The phenomena of spontaneous gravitational symmetry breaking were discussed earlier in \cite{Smolin} where it was shown that an extended Plebanski action of the form of (\ref{action0}), for a gauge group $G$ which contains the Lorentz group, $SO(3,1)$, suffers spontaneous symmetry breaking to an Einstein-Yang-Mills theory with a Yang-Mills gauge group in $G/SO(3,1)$.  The same phenomena were demonstrated by Torres-Gomez and Krasnov 
for the chiral $SU(2)_L$ subgroup of the Lorentz group \cite{kirill-breaking}.  Krasnov also had earlier originated the notion of extending the 
Plebanski action in \cite{kirill-extended}, with $G$ taken to be the chiral left handed space-time connection valued in $SU(2)_L$.  He has also explored a closely related set of theories whose actions are purely functions of connections, and demonstrated the phenomena of gravitational spontaneous symmetry breaking there \cite{kirill-connection}.

In \cite{Nesti:2007jz}, within the framework of left and right-handed gravi-weak unification models, namely $SL(4,\mathbb{C})_L \times SL(4,\mathbb{C})_R$ and its extension $GL(4,\mathbb{C})_L \times GL(4,\mathbb{C})_R$, Nesti has studied a parity breaking coupling of gravitons with combination of opposite helicities to matter. Nesti and Percacci have discussed issues related to the Higgs phenomenon and the electro-weak symmetry breaking in \cite{Nesti:2007ka}, and elaborated those topics for the gravi-GUT unification model they have presented in \cite{Nesti:2009kk}. The latter work develops a different perspective that the one addressed in \cite{Nesti:2007jz}, in that the gravi-weak and color gauge sectors have been accounted separately in \cite{Nesti:2007jz}.

To discuss the dynamics in more detail, as well as to specify the modified reality conditions it is convenient to change to two component spinor indices \cite{Penrose:1985jw, TeWo}. $A,B=0,1$ are left handed spinor indices while $A',B'=0',1'$ are right handed spinor indices. This allows us to easily distinguish the left and right handed fields.   The connection decomposes into
\f
A^{ab}=A^{AA' BB'} = \varepsilon^{AB} A^{A'B'} + A^{AB} \varepsilon^{A'B'}
\ff
and the two forms $B^{ab}$ similarly decompose.  The scalar fields $\Psi_{abcd}$ decompose into pure spin two fields represented by
$\Psi_{ABCD}$ and $\Psi_{A'B'C'D'}$, both totally symmetric, and mixed components $\Psi_{ABA'B'}$ on symmetric pairs of indices.  Thus, 
\f
\Psi_{ABCD}= \Psi_{(ABCD)}
\ff
and the same for primed indices represents the spin two field.

The action now takes the form,
\begin{eqnarray}
S& = & \int \frac{\imath}{4 \pi G} \left \{  B^{AB} \wedge F_{AB} -  B^{A'B'} \wedge F_{A'B'}  + \frac{\lambda}{6G} (B_{AB}\wedge B^{AB} - B_{A'B'}\wedge B^{A'B'} ) \right.
 \nonumber \\
 &&\left. - \frac{1}{2} \Psi_{ABCD} B^{(AB} \wedge B^{CD)}
 + \frac{1}{2} \Psi_{A'B'C'D'} B^{(A'B'} \wedge B^{C'D')}
 - \Psi_{A'B'AB} B^{A'B'} \wedge B^{AB} \right \}
 \nonumber \\
 &&+\frac{\imath g^2}{2} (  \Psi_{ABCD}^2 +    \Psi_{A'B'C'D'}^2 + \Psi_{ABA'B'}^2   ) (B_{AB}\wedge B^{AB} - B_{A'B'}\wedge B^{A'B'} )\,,
 \label{action1}
\end{eqnarray}
where $\lambda = G\Lambda $ is the dimensionless cosmological constant.   

To describe the real world we have to impose reality conditions, which restrict the solutions of the theory to those in which the metric is real.
This can be done directly, in spite of the fact that the metric is not a fundamental field in the action.  Instead, we make use of the remarkable fact that a densitized metric can be constructed which is cubic in the $B$ fields.  In fact, two metrics can be built, out of the left and right parts of
$B$, which we call the left and right  Urbantke metrics \cite{Urbantke, CDJM91} 
\f
\tilde{g}^L_{\mu \nu} = \varepsilon^{\gamma \delta \rho \sigma} B_{\mu \gamma A }^{B} B_{\nu \delta C}^{A} B_{\rho \sigma B}^{C}\,,   
\ff
\f
\tilde{g}^R_{\mu \nu} = \varepsilon^{\gamma \delta \rho \sigma} B_{\mu \gamma A' }^{B'} B_{\nu \delta C'}^{A'} B_{\rho \sigma B'}^{C'} \,, 
\ff
in which $ \varepsilon^{\alpha \beta \gamma \delta} $ is the Levi-Civita symbol and over tildes label tensor densities of weight $-1$, {\it i.e.} tensor densities transforming like a covariant tensor times $\sqrt{-g}$, $g_{\mu\nu}$ being the space-time metric. Note that in the symmetric solution these are equal to each other while  in the asymmetric solution they differ.   The reality conditions we propose are that both left and right  handed Urbantke metrics are real.  \\

To summarize, we make four physical hypotheses:

\begin{itemize}

\item{}The $SU(2)$ of the weak interactions is unified with the chiral representation of gravity in a single $SL(2, \mathbb{C})$ connection.  This was proposed earlier by Alexander \cite{Alexander:2007mt}  and by Nesti and Percacci \cite{Nesti:2007ka}. A toy-model in $3D$ was presented in \cite{Alexander:2011jf}  by Alexander, Marcian\`o and Tacchi, together with its spin-foam quantization.

\item{}The chirality of the standard model arises from a spontaneous breaking of parity in the gravitational dynamics.  It is the weak
interactions that break parity because the weak $SU(2)$ gauge connection is in fact a chiral half of what is originally the space-time connection.

\item{}This mechanism also explains why parity is maximally violated in the weak interactions.  The parity mirror of the coupling of weak isospin to matter is
the coupling of the left handed part of the space-time connection to left-handed spinors.

\item{}Under the symmetry breaking, right handed space-time spinors become internal isospinors.  
More specifically, consider the Higgs field, a space-time scalar valued in the $\frac{1}{2}$ of gauged isospin and the sterile neutrino (or right handed neutrinos in general) which are isospin singlets but space-time spinors.  These are mirrors of each other under the parity symmetry that exchanges
the $SU(2)_L$ and $SU(2)_R$ parts of the original connection, and are hence unified in a single Dirac spinor.

\end{itemize}

The basic dynamics of the $SL(2,\mathbb{C})_\mathbb{C}$ extended Plebanski action are detailed in the next section. Sections 3 and 4 describe the symmetric and broken phases of solutions.  The imposition of reality conditions is discussed in sections 5. For the theory to truly unify the electroweak interactions with gravity there must be a $U(1)$ in the theory. This can be incorporated most
simply by extending $SL(2,\mathbb{C})_R$ to $GL(2,\mathbb{C})_R$ as is discussed in section 6.  But if we keep the philosophy that parity is only broken spontaneously, there must be another $U(1)$ gauge field coming from extending $SL(2,\mathbb{C})_L$ to $GL(2,\mathbb{C})_L$. Matter coupling is discussed in section 7, and some possible phenomenological consequences are spelled out in the conclusions, in section 8. Finally, the appendix contains a summary of the Infeld-Van der Waerden map, in section A, and of the conventions and recurrent identities we have been making use of, in section B.

\section{Field equations}

We now exhibit the  field equations.  Because the reality conditions are subtle we start with the complexification of the theory and study phase invariant reality conditions below.\\

\noindent 
We write the equations of motion: from variation with respect to the $B^{AB}$ and $B^{A'B'}$ fields we obtain
\begin{eqnarray}
F_{AB} &=& \Psi_{ABCD} \ B^{CD} +\Psi_{ABA'B'} \ B^{A'B'} 
- \left(\frac{\lambda}{3G} + 4\pi G g^2 \Psi^2 \right) B_{AB}\,,
\label{Beom}
 \\
F_{A'B'} &=& \Psi_{A'B'C'D'} \ B^{C'D'} -\Psi_{A'B'AB} \ B^{AB} 
+\left(- \frac{\lambda}{3G} +4 \pi G g^2 \Psi^2 \right) \ B_{A'B'}\,,
\label{B'eom}
\end{eqnarray}
whilst varying with respect to the multiplet of scale fields we find
\begin{eqnarray}
\Psi_{ABCD} = \frac{1}{8 \pi G g^2 W} \ B_{(AB} \wedge B_{CD)} \,,
\label{psieom}
\\
\Psi_{A'B'C'D'} = -\frac{1}{8 \pi G g^2 W} \ B_{(A'B'} \wedge B_{C'D')} \,,
\label{psi'eom}
\\
 \Psi_{ABA'B'} = \frac{1}{4  \pi G g^2 W} \ B_{AB}  \wedge B_{A'B'} \,,
 \label{psimixedeom}
\end{eqnarray}
where
\f
W= B_{AB}\wedge B^{AB} - B_{A'B'}\wedge B^{A'B'} \,.
\ff
Finally, variation with respect to the connection components gives
\f
{\cal D} \wedge B_{AB} = {\cal D}' \wedge B_{A'B'}=0 \,, 
\ff
in which ${\cal D}$ stands for the covariant derivative with respect to $A^{AB}$, whilst ${\cal D}'$ stands for the covariant derivative with respect to $A^{A'B'}$. 

\section{Symmetric solution}

We begin with a left-right symmetric solution of the theory.  
We expand the $B^{AB}$ and $B^{A'B'}$ in $g$
\f
B_{AB}= B^{(0)}_{AB} + g^2  b_{AB} \,,
\label{Bshift}
\ff
\f
B_{A'B'}= B^{(0)}_{A'B'} + g^2 b_{A'B'} \,.
\ff
We then solve the equations of motion (\ref{psieom}) and (\ref{psi'eom}) order
by order in $g$.  
We have to leading order on the left side
\f
B^{(0)}_{(AB} \wedge B^{(0)}_{CD)} =0\,,
\label{simplicity1}
\ff
while to order $g^2$
\f
b_{(AB} \wedge  B^{(0)}_{CD)} +\frac{g^2}{2}b_{(AB} \wedge b_{CD)} = 4 \pi G \Psi_{ABCD} W. 
\ff
There then must exists  frame field $e^{AA'}$ such that
\f
B^{(0)}_{AB} = e_A^{\,\,A'} \wedge e_{BA'} =\Sigma_{AB} \,.
\ff
Similarly on the right side we have the same equations of motion, 
\f
B^{(0)}_{(A'B'} \wedge B^{(0)}_{C'D')} =0
\ff
and 
\f
B^{(0)}_{(A'B'} \wedge b_{C'D')} +\frac{g^2}{2}b_{(A'B'} \wedge b_{C'D')} =-4 \pi G  \Psi_{A'B'C'D'} W \,,
\ff
which tells us that there must exist a second frame field $f^{AA'}$ such that
\f
B^{(0)}_{A'B')} = f_{\,\,A'}^{A} \wedge f_{A B'} \,\chi' =\Sigma_{A'B'}^\prime (f)\, \chi' \,.
\ff
The two frame fields, $e^{AA'}$ and $f^{AA'}$ are coupled through (\ref{psimixedeom})
which to leading order give
\f
\Sigma^{AB} (e) \wedge \Sigma^{' A'B'} (f) =0\,.
\ff
This is solved by
\f
f^{AA'} = h \ e^{AA'}\,,
\ff
where $h$ is a function. Speziale shows that in general the symmetric solutions give a bi-metric theory with eight degrees of freedom \cite{Speziale:2010cf}. 

\section{Symmetry breaking solution}

We keep the above solution on the left, unprimed, side, so we continue to expand as in (\ref{Bshift}).
Thus, we have a frame field $e^{AA'}$ from the solution to (\ref{simplicity1}).

However on the right handed side we do something else.  
We image that the equation of motion for $F_{A'B'}$, eq. (\ref{B'eom}) is dominated by the term $(B_{A'B'} \lambda)/3G$ and by the term $B^{CD} (B_{A'B'} \wedge B_{CD})/(4 \pi G g^2 W)$. In doing so, we have assumed $B_{A'B'}$ to be order $g^2$ and higher and we have imaged the scaling $\lambda \, g^2 = \xi$, with $\xi$ a fixed dimensionless real parameter, so that
\f \label{rightshift}
F_{A'B'} \approx -  B^{CD}  \frac{ B_{( A'B'} \wedge B_{CD)}  }{ 4 \pi G g^2 W}  - \frac{\lambda}{3G} B_{A'B'} + O(g^2)\,.
%\dots \,.
%+ \frac{g^2}{G} \, b^{A'B'}  + O(g^3)   \,.
\ff
We may now expand the left handed $B$ fields as $B^{AB}=\Sigma^{AB} + O (g^2)$, which to zeroth order leads to
\f
W= 24\, \imath \,e + O(g^2) \,,
\ff
and invert (\ref{rightshift}) in order to obtain an expression for $B^{A'B'}$ in terms of $F^{A'B'}$ and its dual. We then realize the relation between $B^{A'B'}$ and $F^{A'B'}$ shifting the $B^{A'B'}$ field by
\f \label{shift}
B_{A'B'}= - \pi G g^2  \left(  \delta_\xi 1\!\! 1 + \gamma_\xi\,  \star \right) \ F_{A'B'} + g^6 b_{A'B'}\,,
\ff
where $\star$ stands for the space-time Hodge dual with indices suppressed, $1\!\!1$ acts as the identity operator on 2-forms, and
\f \label{adri}
\delta_\xi=  \left( \frac{1}{16} + \frac{\xi}{3}  \right)   \frac{1}{\left(  \left( \frac{1}{16} + \frac{\xi}{3}  \right) ^2 - \left(\frac{3}{128}\right)^2\right)}\qquad {\rm and } \qquad  \gamma_\xi=-\frac{3\, \delta_\xi}{128  \left( \frac{1}{16} + \frac{\xi}{3}  \right)}\,.
\ff

We can check that the shifting term in (\ref{shift}) is small in solutions to equations of motion and compute that $b_{A'B'}$ is suppressed in power of $G$:
\f
b_{A'B'}\!=\! -\pi^4 G^3 \! \left( 
(\delta^2_\xi+\gamma^2_\xi) 1\!\! 1 + 2 \delta_\xi \, \gamma_\xi \star \right)\! F^{C'D'} \! \! \left(  \delta_\xi 1\!\! 1 + \gamma_\xi \star \right) F_{A'B'} \!\wedge\! \left(  \delta_\xi 1\!\! 1 + \gamma_\xi \star \right) F_{C'D'}  \!+\! O(g^6) \,.
\ff

To understand the effect of this shift we solve the equations of motion for the
$\Psi$ multiplet, (\ref{psieom})-%,\ref{psi'eom},
(\ref{psimixedeom}), and plug the result back into the action to find
\begin{eqnarray}
S& = & \int \frac{\imath}{4 \pi G} \left \{  B^{AB} \wedge F_{AB} -  B^{A'B'} \wedge F_{A'B'} +\frac{\lambda}{6G} (B_{AB}\wedge B^{AB} - B_{A'B'}\wedge B^{A'B'} ) \right \}
 \nonumber \\
  &&+\frac{ 81\, \imath}{128 \pi^2 G^2 g^2 W} \Big(   (B_{AB}\wedge B_{CD})^2 + (B_{A'B'}\wedge B_{C'D'} )^2 
 -4 (B_{AB}\wedge B_{A'B'})^2
 \Big) \,.
 \label{action2}
\end{eqnarray}

We incorporate the shift (\ref{rightshift}) together with
\f
B^{AB} = \Sigma^{AB} + g^2 b^{AB}
\ff
to write the action as
\f
S=S^{(0)} (e^{AA'}, A_{AB}, A_{A'B'} ) + S^{(1)} (b_{AB}, b_{A'B'}, e^{AA'}, A_{AB}, A_{A'B'} )\,,
\ff
where the leading order action $S^{(0)}$ is 
\begin{eqnarray}
S^{(0)} & = & \int \frac{\imath }{4\pi G} \, \Sigma^{AB} \wedge F_{AB} 
+ \frac{\lambda}{12 \pi G^2} \, e \nonumber \\
 &&
- \frac{e}{4g^2_{YM}} F^{A'B'}_{\mu \nu } F_{A'B' \rho \sigma} g^{\mu\rho} g^{\nu \sigma} 
- \imath \, \Theta \, F^{A'B'} \wedge  F_{A'B' } 
\nonumber \\
&&
+ \frac{9 G^2 }{(16 \pi)^2 \lambda^2  e}\, (F_{(A'B'}\wedge F_{C'D') } )^2
\label{action3}\,.
\end{eqnarray}
In this latter expression the Yang-Mills coupling constant is

\f
-\frac{1}{4 g^2_{YM}}= g^2 \left[ \delta_\xi\, \gamma_\xi \left( \xi \frac{\pi^2}{3} - \frac{1}{64} -74 \right) + \gamma_\xi \right]\,,
\label{gYM}
\ff
while the $\Theta$ angle is
 \f
\Theta= g^2 \left[ (\delta^2_\xi + \gamma^2_\xi) \left( \xi \frac{\pi^2}{6} - \frac{1}{128} -37 \right) + \delta_\xi \right]\,.
 \ff
Notice that for $\xi\sim 10^{-1}$ or smaller we would get $g^2 \, g_{YM}^2\sim 10^{-4}$. 

The $b^{AB}$ and $b^{A'B'}$ are auxiliary fields which are 
determined by variation of the higher order action $S^{(1)}$, namely
 \begin{eqnarray}
\!\!\!\!S^{(1)}& \!=\!& \int \frac{\imath g^2}{4 \pi G} \left \{  
 b^{AB} \wedge F_{AB} +\frac{\lambda }{3G} b_{AB}\wedge \Sigma^{AB} +\frac{\lambda g^2 }{3G}  b_{AB}\wedge b^{AB} 
 + \frac{g^2  \lambda}{3G} b_{A'B'}\wedge b^{A'B'} 
 \right \}
 \nonumber \\
 &&+\frac{81\, \imath}{128 \pi^2 G^2 g^2 \, W} \Big(  (B_{AB}\wedge B_{CD})^2 + (B_{A'B'}\wedge B_{C'D'} )^2 - 4 (B_{AB}\wedge B_{A'B'})^2
 \Big)^{(1)}\!\!\!,
 \label{action4}
\end{eqnarray}
 where by the last "$(É)^{(1)}$" we mean that the zeroth order terms present in $S^{(0)}$ in eq. (\ref{action3}) are absent. 
 
The action $S^{(1)}$ is a quartic polynomial, non-derivative in $b^{AB}$ and $b^{A'B'}$. These latter fields are then determined by the solution of local, non-derivative cubic equations. By solving these equations we get higher order interactions in the physical fields, $e^{AA'}, A^{AB}$ and $A^{A'B'}$.

\section{Reality conditions}

A crucial part of this construction is a modified form of the reality conditions. Initially we regard all fields as complex (for the lorentzian case), and then specify reality conditions which  are to be imposed on the solutions of the equations of motion. 

We first review the standard reality conditions imposed in the Ashtekar formulation, and then introduce our new proposal.

The standard reality conditions are to take the three metric as real,
\f
{\tilde{q}}^{ab \ *}=\tilde{q}^{ab }\,.
\ff
while the connection satisfies the non-linear condition.
\f \label{co1}
A_a^{(L)i} + A_a^{(R)i} = 2 \Gamma (e)_a^i \,,
\ff
Here we indicate the left and right connection by
\f
A_a^{AB}= A_a^{(L)i} \sigma^{AB}_i \,, \ \ \ \ \  A_a^{A'B'}= A_a^{(R)i}\sigma^{A'B'}_i \,,
\ff
where the $i$ index labels the three Pauli matrices $\sigma_i$. 

These bind the left and the right parts of the connection and so prevent the theory from existing in the parity broken phase.
In that asymmetric phase, we might impose different reality conditions:
\f \label{co2}
(A_a^{(L)i} )^* + A_a^{(L)i}  = 2 \Gamma (e)_a^i    , \ \ \qquad\ \   (A_a^{(R)i})^*= A_a^{(R)i} .
\ff

However the reality conditions are part of the definition of the theory.  They determine the inner product of the quantum theory.
If the symmetry breaking is to be dynamical we do not want to impose different reality conditions on different phases of the theory.
We want instead a single set of reality conditions that governs the whole theory.  We can do this the following way:

We differentiate the right and left two forms as
\f
B^{A'B'} = B^{L\, i}\, \sigma_i^{A'B'} , \ \ \ \  B^{AB} = B^{R\, i} \,\sigma_i^{AB}\,.
\ff
We then use these to define the left and right Urbantke metrics \cite{Urbantke}:
\f
\tilde{g}^R_{ab} = B_{ac}^{R i} B_{bd}^{R j} B_{ef}^{R k}   \varepsilon_{ijk} \epsilon^{bdef}\,,
\ff
\f
\tilde{g}^L_{ab} = B_{ac}^{L i} B_{bd}^{L j} B_{ef}^{L k}   \varepsilon_{ijk} \epsilon^{bdef}\,.
\ff
Note that in the symmetric solution
\f
\tilde{g}^L_{ab} = {\rm det}(e) \, e_a^{AA'} e_{b \ \ A'}^B\,,
\label{gL}
\ff
while on the right
\f
\tilde{g}^R_{ab} = {\rm det} (f)\, f_a^{AA'} f_{b \ \ A'}^B\,.
\label{gR}
\ff

In the asymmetric solution (\ref{gL}) holds but instead of (\ref{gR}) we have
a cubic in the Yang-Mills field strength
\f
\tilde{g}^R_{ab} = - 27 \frac{G^3}{\lambda^3} \ F_{ac}^{ i} F_{bd}^{ j} F_{ef}^{ k}   \epsilon_{ijk} \varepsilon^{bdef}\,.
\ff

In either case the correct reality conditions are
\f
\tilde{g}^L_{ab} = (\tilde{g}^L_{ab})^*\,,
\ff
\f
\tilde{g}^R_{ab} = (\tilde{g}^R_{ab})^*\,.
\ff
In the symmetric case this tells us that both left and right handed metrics are real, whereas in the asymmetric solution we learn that $\tilde{g}^R_{ab}$ is real and the Yang-Mills connection $\omega_a^i$ is real and hence in $SU(2)$.

These can be implemented by adding these reality conditions to the action so they become equations of motion which arise by varying new Lagrange multipliers $\lambda^{ab}_{L,R}$:  
\begin{eqnarray}
S^{w rc} & = & \int \frac{\imath}{4 \pi G} \left \{  B^{AB} \wedge F_{AB} -  B^{A'B'} \wedge F_{A'B'} + \frac{\lambda}{6G} (B_{AB}\wedge B^{AB} - B_{A'B'}\wedge B^{A'B'} ) \right.
 \nonumber \\
 &&\left. - \frac{1}{2} \Psi_{ABCD} B^{(AB} \wedge B^{CD)}
 + \frac{1}{2} \Psi_{A'B'C'D'} B^{(A'B'} \wedge B^{C'D')}
 - \Psi_{A'B'AB} B^{A'B'} \wedge B^{AB} \right \}
 \nonumber \\
 &&+\frac{\imath g^2}{2} (  \Psi_{ABCD}^2 +    \Psi_{A'B'C'D'}^2 + \Psi_{ABA'B'}^2   ) (B_{AB}\wedge B^{AB} - B_{A'B'}\wedge B^{A'B'} )
  \nonumber \\
  &&+ \lambda^{ab}_R \Big(\tilde{g}^R_{ab} - (\tilde{g}^R_{ab})^* \Big)
  + \lambda^{ab}_L \Big(\tilde{g}^L_{ab} - (\tilde{g}^L_{ab})^* \Big)\,.
\end{eqnarray}

The $B$ equation of motion (\ref{Beom}) is modified by
\f
F_{AB} = \Psi_{ABCD} \,B^{CD} +\Psi_{ABA'B'} \,B^{A'B'} 
-(\frac{\lambda}{3G} + 4\pi G g^2 \Psi^2 ) \,B_{AB}
+ 4\pi \imath \,G\, \lambda^{ef}_R\, \frac{\delta \tilde{g}^R_{ef}}{\delta B^{AB}}\,.
\ff
But the new term vanishes because the equation of motion for $B^{AB *}$ yields
\f
 \lambda^{ef}_R \frac{\delta \tilde{g}^{R *}_{ef}}{\delta B^{AB *}}=0\,,
\ff
which implies that $\lambda^{ab}_R$ vanishes.  Meanwhile, variation of $\lambda^{ab}_R$ enforces the reality of $\tilde{g}^{R}_{ef}$.

\section{Adding $U(1)$ factors: photons}

We can incorporate electro-weak unification by adding a $U(1)$ factor\footnote{The inclusion of a $U(1)$ gauge field by extending the
Plebanski action was also studied in \cite{kirill-U(1)}.}.
  This is done most naturally by extending the $SL(2,\mathbb{C})_L$ gauge symmetry to\footnote{A subgroup of $GL(2,\mathbb{C})_L$ is $SL(2,\mathbb{C})_L \times U(1)_L$.} $GL(2,\mathbb{C})_L$, and similarly for the right component gauge group. The gauge fields $A^{A'B'}$ are then no longer symmetric in $AB$.  
\f
A^{A'B'}= A^{(A'B')} +\varepsilon^{A'B'} a'
\ff
defining the $U(1)$ gauge field $a'$.
If we want to continue to follow our hypothesis that left-right breaking occurs only spontaneously we should do this on the left as well,
so
\f
A^{AB}= A^{(AB)} +\varepsilon^{AB} a\,.
\ff
We use the same action (\ref{action1}), which becomes the previous action plus a $U(1)_\mathbb{C}$ factor.
\f
S=S^{SL(2,\mathbb{C})_\mathbb{C}} + S^{U(1)_\mathbb{C}}
\ff
where $S^{SL(2,\mathbb{C})_\mathbb{C}}$ is the previous action (\ref{action1}),  (with $W$ extended as below) and
\begin{eqnarray}
S^{U(1)_\mathbb{C}}& = & \int \frac{\imath }{4\pi G} \left \{ B \wedge f  - B' \wedge f' 
 +\frac{\lambda}{6G} (B \wedge B - B' \wedge B' ) \right.
 \nonumber \\
 && \left.
 - \frac{1}{2} \Psi_{\cdot \cdot} B \wedge B - \frac{1}{2} \Psi_{\cdot' \cdot'} B' \wedge B'
 - \frac{1}{2} \Psi_{\cdot \cdot'} B \wedge B' \right.
  \nonumber \\
 && \left.
 -  \Psi_{AB \cdot} B^{AB} \wedge B  - \Psi_{A'B' \cdot} B^{A'B'} \wedge B
  -  \Psi_{AB \cdot'} B^{AB} \wedge B' -  \Psi_{A'B' \cdot'} B^{A'B'} \wedge B' \right \}
 \nonumber \\
 &&
 +\frac{ \imath g^2}{2} (  \Psi_{\cdot \cdot }^2 +    \Psi_{\cdot' \cdot' }^2 + \Psi_{\cdot \cdot' }^2 +    \Psi_{\cdot' AB }^2 + \Psi_{\cdot' A'B'}^2 + \Psi_{AB \cdot}^2 + \Psi_{\cdot A'B'}^2) \, W\,,
 \label{actionU(1)}
\end{eqnarray}
where $B$ and $B'$ are abelian two-forms and $f=da$, $ f'=da'$ and, now,
\f
W= B_{AB} \wedge B^{AB}  - B_{A'B'} \wedge B^{A'B'} +B \wedge B -  B' \wedge B'\,.
\ff
We can again solve for the $\Psi$ equations of motion to cast the action in the form
\begin{eqnarray}
\!\!\!\!\!\!S^{U(1)_\mathbb{C}}& \!=\! & \int \frac{\imath }{4\pi G} \left \{ B \wedge f  - B' \wedge f' 
 +\frac{\lambda}{6G} (B \wedge B - B' \wedge B' ) \right \}  +
 \nonumber \\
% && \left. - \frac{1}{2} \Psi_{\cdot \cdot} B \wedge B - \frac{1}{2} \Psi_{\cdot' \cdot'} B' \wedge B' - \frac{1}{2} \Psi_{\cdot \cdot'} B \wedge B' \right. \nonumber \\
% && \left. -  \Psi_{AB \cdot} B^{AB} \wedge B  - \Psi_{A'B' \cdot} B^{A'B'} \wedge B   -  \Psi_{AB \cdot'} B^{AB} \wedge B' -  \Psi_{A'B' \cdot'} B^{A'B'} \wedge B' \right \}
% \nonumber \\
&&
+\frac{81\, \imath}{128 \pi^2 G^2 g^2 }  \Big(
(B \wedge B)^2 + (B' \wedge B')^2 + (B \wedge B')^2 - 4 (B \wedge B_{AB})^2 +  \nonumber\\
&& 
-4 (B \wedge B_{A'B'})^2 - 4 (B' \wedge B_{AB})^2 -4  (B' \wedge B_{A'B'})^2
\Big).
 \label{actionU(1)2}
\end{eqnarray}
Note that if $g=0$ the $\Psi_{\cdot \cdot}$ equation of motion gives
\f
B \wedge B =0\,,
\ff
which implies $B=0$. So this extension of the gauge group is only possible in the extended (as opposed to the unextended) Plebanski action.  We then have no choice but to shift the $U(1)$ fields:
\f
B =- \pi G g^2  \left(  \delta_\xi 1\!\! 1 + \gamma_\xi \, \star \right) \, f + g^6 \, b \,,  \qquad \ \ \ \ B' =- \pi G g^2  \left(  \delta_\xi 1\!\! 1 + \gamma_\xi \, \star \right) \, f' + g^6\,  b' \,,
\label{rightshiftU1}
\ff
where $\delta_\xi$ and $\gamma_\xi$ have been defined in (\ref{adri}).\\

Again this gives a zeroth order action plus an action for the auxiliary fields, $b$ and $b'$.  Below we only write the $U(1)_\mathbb{C}$ part of the leading order action:
\begin{eqnarray}
S_{(0)}^{U(1)_\mathbb{C}} &=&
 \frac{e}{4g^2_{YM}} (f_{\mu \nu } f_{\rho \sigma}  +f'_{\mu \nu } f'_{\rho \sigma}   )g^{\mu\rho} g^{\nu \sigma} 
+  \Theta ( f \wedge  f + f' \wedge f' ) +
\nonumber \\
&&
+ \frac{9\, g^2\, G^2 }{256\, \xi^2  e} \, \Big(  (\tilde{f} \wedge \tilde{f} )^2 + (\tilde{f}' \wedge \tilde{f}' )^2 +  (\tilde{f} \wedge \tilde{f}' )^2 - 4  (\tilde{f} \wedge F_{AB}  )^2 +  
\nonumber\\
&&
- 4  (\tilde{f} \wedge F_{A'B'} )^2 -4  (\tilde{f}' \wedge F_{AB} )^2 - 4  (\tilde{f}' \wedge F_{A'B'} )^2 \Big)\,,
\label{actionU(1)3}
\end{eqnarray}
where we have used the shorthand notations $\tilde{f} =  \left(  \delta_\xi 1\!\! 1 + \gamma_\xi \, \star \right) f$ and $\tilde{f}' =  \left(  \delta_\xi 1\!\! 1 + \gamma_\xi \, \star \right) f'$. We see the following interesting features:
\begin{itemize}

\item{} the two $U(1)$ factors have the same Yang-Mills coupling constant as the $SU(2)_L$ factor, so there is coupling constant unification;

\item{}however they will couple differently to matter as we will see;

\item{}there is a universal four point coupling of vector potentials of the form $(F \wedge F)^2$ which has a universal coupling
\f
\lambda_{4 -{\rm point}}= \frac{9 \,g^2\, G^2 }{256\,  \xi^2 }\, \sim  \frac{9 \, g^2 }{256\, M_p^4  \xi^2 }\
\ff
which is quite small.  

\end{itemize}

\section{Matter couplings}

Matter couplings are tricky to write because there is no metric or frame field initially (as of instance in \cite{Jacobson:1988qta}), but only a $B^{AB}$ field.  Couplings to scalars and additional gauge fields can be done through the Urbantke metric, but they involves non-polynomial couplings.   The simplest coupling is to chiral spinors, with the following action \cite{CDJM91}:
\f
S^{\rm Dirac}_L = \int B^{AB} \wedge \rho_A \wedge ({\cal D} \lambda )_B + \tau_{ABC} \wedge B^{(AB} \wedge \rho^{C)}\,.
\ff
This works like the Plebanski actions above, but here $\tau^{ABC} = \tau^{(ABC)}$ is a Lagrange multiplier 1-form whose variation, together with the leading order solution for $B^{AB}$, yields
\f
\Sigma^{(AB} \wedge \rho^{C)} =0\,,
\label{taueom}
\ff
which is solved by inventing a complex conjugate spinor, $\bar{\lambda}_{A'}$ such that
\f
\rho^A = e^{AA'} \bar{\lambda}_{A'}\,.
\ff
Putting this back into the action we find an effective action
\f
S^{\rm Dirac}_L = \int \bar{\lambda}_{A'}  e^{A'}_A  \wedge \Sigma^{AB}  \wedge ({\cal D} \lambda )_B \,,
\ff
which yields the Weyl theory for a right handed spinor $\lambda^A$.
\\

Let us now consider the right handed side.  By symmetry, we must start the same way:
\f
S^{\rm Dirac}_R = \int B^{A'B'} \wedge \rho_A' \wedge ({\cal D} \lambda )_B' + \tau_{A'B'C'} \wedge B^{(A'B'} \wedge \rho^{C')}\,.
\ff
Note that there is no relation between $\lambda^{A'}$ and $\bar{\lambda}^{A'}$, indeed the latter is not even a field in the fundamental action.

In the symmetric solution things work the same way on the right side as the left, and the result is that the fields combine to make a Dirac spinor.
But on the symmetry breaking side things on the right side are not so simple.  Instead of (\ref{taueom}) we have
\f
F^{(A'B'} \wedge \rho^{C')} =0\,,
\label{tau'eom}
\ff
which does not have any simple general solution.  

In terms of its transformation properties, in the symmetry broken phase, $\lambda_{A'}$ is a space-time scalar and
weak spinor, so it has the quantum numbers of the Higgs boson.  

\section{Conclusion}

\noindent 

Ever since the discovery and experimental success of the standard electroweak theory, the origin of the weak interaction's chirality has remained a mystery. In this work we we have reached the conclusion that a parity symmetric theory of gravity holds the key to the chiral origin and maximal parity violation of the weak interaction. In particular, we describe a parity symmetric theory of gravity that has a symmetry broken phase, which organizes the degrees of freedom to give rise to general relativity coupled to a $SU(2)$ Yang-Mills theory.   The emergence of gravity and the weak interaction is made possible because gravity has been shown to be completely described in terms of purely left-handed variables \cite{Jacobson:1987yw}.   This leaves the right handed connection to function as the weak interaction connection.

One concern is that the expansion in which we understand the symmetry broken phase involves small $g$ and large $\lambda$ --- the dimensionless cosmological constant. (See (\ref{rightshift}) and (\ref{gYM}).)  Since $\lambda$ is the bare cosmological constant, it might be possible to imagine that it must be large to cancel contributions coming from radiative corrections and symmetry breaking, but this will require more investigation.

Extending the $SL(2, \mathbb{C})_\mathbb{C}$ symmetry group to $GL(2, \mathbb{C})_\mathbb{C}$, enables us to account for two additional $U(1)$ sectors, one of them describing the $U(1)_{Y}$ and the other accounting for an extra abelian gauge group that we can speculate may be eventually related to dark matter.  We note that the theory we have discussed naturally supplies an extra $U(1)$ which has been suggested both as a constituent of dark matter \cite{Lang, Ackerman:2008gi} and as possibly relevant to the di-photon excess seen at the LHC \cite{aad,cha}. In our model, the Higgs boson arises with the correct quantum numbers and in the symmetry restored gravitational theory is identified with a sterile neutrino under a party transformation.  It is important to see any effects of the symmetry breaking in the neutrino and Higgs sector, such as new interaction vertices; we leave this question to future work.  In this work, we did not provide a mass generation's mechanism, such as spontaneous symmetry breaking (SSB), although it is not difficult to implement this into our model.  We will address the issue of SSB in a forthcoming paper \cite{SMA2}, together with the role of the extra $U(1)$ gauge sector, in order to unveil its consequences for current and upcoming LHC experiments.

\appendix

\section{From Lorentz indices to spinorial indices: Infeld-Van der Waerden map}

\noindent We started from a theory that is Lorentz invariant, whose ``objects'' in the starting action are Lorentz algebra valued tensor fields. We have then recast the action in terms of spinor fields. If we want to account for the usual Plebanski theory is spinorial variables, we may start the analysis considering in stead of a $SO(3,1)$ principal fiber bundle an $SL(2,\mathbb{C})$ principal fiber bundle, where $SL(2,\mathbb{C})$ is the universal covering of group of $SO(3,1)$. Lorentz tensor fields are then no more simply sections of $\mathcal{P}_{SL(2, \mathbb{C})}$, but they are allowed to be multi linear maps, or eventually anti-linear maps if we use complex conjugation. Sections of these bundles are called spinor fields. We are then lead to the construction of a spinor algebra, that is mapped in the Lorentz algebra valued tensor fields algebra through the Infeld-van der Waerden symbols $\sigma$:
\be
T^{r \dots}_{ \ \ \ s \dots} \rightarrow T^{A A' \dots }_{\ \ \ \ \ \ \ BB' \dots}= \sigma^{AA'}_r \dots \sigma^s_{BB'} \dots T^{r \dots}_{ \ \ \ s \dots}\,,
\ee  
where again the pair of capital latin indices $A,A'$ take the values $0,1$. We identify $\sigma^{AA'}_0$ with the $2 \times 2$ unit matrix and $\sigma^{AA'}_1, \ \sigma^{AA'}_2, \ \sigma^{AA'}_3$ with the Pauli matrices, satisfying the relations
\be
\sigma^{AA'}_r\ \sigma_{BB'}^r= \delta^A_{\ B}\, \delta^{A'}_{\ B'}\,, \qquad \sigma^{AA'}_r\ \sigma_{AA'}^s= \delta^r_{\ s}\,.
\ee
This map allows us to recover spinor fields from Lorentz algebra tensors, and thus to rewrite our starting actions for the $SL(2,\mathbb{C})_\mathbb{C}$ and $GL(2,\mathbb{C})_\mathbb{C}$ symmetric theories. 

We start here below reviewing the spinor equivalent of the objects entering our action. Tetrads are recast using the map in $e^a \rightarrow e^{AA'}=e^a \, \sigma^{AA'}_a$. Writing the Lorentz connection components requires to select a convention on the order the primed and unprimed indices appear: what matters is indeed the relative order of the primed with respect to the umprimed indices associated to the Lorentz index through the Infeld-van der Waerden symbols. Since the Lorentz connection is written in terms of a pair of antisymmetric indices such that $A_{ab}=-A_{ba}$, we have then that $A_{ABA'B'}=-A_{BAB'A'}$, from which it follows that the Lorentz connection can be decomposed in 
\be
A_{ABA'B'}=A_{AB}\  \varepsilon_{A'B'} + A_{A'B'} \ \varepsilon_{AB}\,,
\ee
in which now the components $A_{AB}$ and $ A_{A'B'}$ are symmetric in the swap of indices in order to fulfill the property $A_{ABA'B'}=-A_{ABA'B'}$. 

We use a convention such that $\varepsilon_{AB}=\varepsilon^{AB}$, in which $\varepsilon_{AB}$ is the Levi-Civita symbol $\varepsilon_{01}=-\varepsilon_{10}=-1$. Indices are raised and lowered using the so called ``northwest-southeast'' convention, that means $\omega^P_{\ N}=\varepsilon^{PQ} \, \omega_{QN}$ and $e_A^{\ C'}=e^{PC'} \varepsilon_{PA}$. Moreover
\be
\varepsilon_B^{\ A}=\delta^A_B=- \varepsilon^A_{\ B}\,.
\ee

The one forms $A_{MN}$ are $SL(2,\mathbb{C})$ connection with an associated curvature two form that we may recover from the Ricci tensor $R_{\mu \nu} \rightarrow R_{MNM'N'}$, which can be decomposed in
\be
R_{MNM'N'}=R_{MN} \ \varepsilon_{M'N'} + R_{M'N'} \ \varepsilon_{M N} \,,
\ee
in which $R_{MN}$ and $R_{M'N'}$ are as well symmetric in the swap of indices and are expressed by
\be
R_{MN}= d A_{MN} + A_M^{\ P} \wedge  A_{PN}\,, \qquad R_{M'N'}= d A_{M'N'} + A_{M'}^{\ P'} \wedge  A_{P'N'} \,.
\ee

Following this {\it recipe} we can map any Lorentz tensor field in a spinor field. Finally, we recall that
\be
\varepsilon_{MNPQM'N'P'Q'}=\imath \left( \varepsilon_{MP} \varepsilon_{NQ} \varepsilon_{M'Q'} \varepsilon_{N'P'} - \varepsilon_{MQ} \varepsilon_{NP} \varepsilon_{M'P'} \varepsilon_{N'Q'} \right)\,.
\ee
 
Thanks to this decomposition, we may recast the covariant derivative (with respect to the Lorentz connection) acting for instance on the tetrad field as 
\be
D e^{AA'} = d e^{AA'} + A^A_{\ B} \wedge e^{B A'} + A^{A'}_{\ B'} \wedge e^{A B'}\,,
\ee 
which allows for instance to write the Cartan structure equation for the Einstein-Hilbert action in the vacuum as 
\be
D \, e^{(M}_{\ \ Q'} \wedge e^{N)Q'}=0 \qquad  \longleftrightarrow \qquad d (e^I \wedge e^J) - e^K\wedge e^J \wedge A^I_{\ K} + e^I\wedge e^K \wedge A^J_{\ K} =0\,.
\ee

Notice now that we may switch to the Plebanski formulation of gravity by composing in the only two possible combinations allowed the tetrad field in the resulting (Plebanski) two forms
\be
B^{AB}= e^{AC'} \wedge e^B_{\ \ C'} \, \qquad B^{A'B'}= e^{CA'} \wedge e_C^{\ \ B'}\,,
\ee
which in turn allows to rewrite
\be
e^{AA'}\wedge e^{BB'} = - \frac{1}{2} \left( \varepsilon^{A'B'} \ B^{AB} + \varepsilon^{AB} B^{A'B'}\  \right)\,.
\ee

Using the Infeld-van der Waerden map we can recover that the Plebanski 2-forms satisfy the following geometric properties 
\be
\star B^{AB}= -\imath B^{AB}\, \qquad \star B^{A'B'}= \imath B^{A'B'} \,,
\ee
in which we have used the definition of the space-time Hodge star operator \be
\star (dx^\mu \wedge dx^\nu)=\frac{1}{2\, \sqrt{-g}} \ \varepsilon^{\mu\nu\rho\sigma} g_{\rho \alpha} g_{\sigma \beta}\  dx^\alpha \wedge dx^\beta\,.
\ee
Thus $\star$ acts as  an endomorphism on the Plebanski 2-forms. Notice now that the identity $\star B^{AB} \wedge B^{A'B'}= B^{AB} \wedge \star B^{A'B'} $ implies actually that 
\be
B^{AB} \wedge B^{A'B'} =0\,,
\ee
and that finally
\be
B^{AB} \wedge B_{CD} = 4 \imath \delta^{(A}_C \delta^{B)}_D\, \sqrt{-g} d^4 x\,,
\ee
having recognized that $d^4x:=dx^0 \wedge dx^1 \wedge dx^2 \wedge dx^3$. 

Notice that in terms of these Plebanski variables the Einstein Cartan action in presence of a cosmological constant $\Lambda=\lambda/G$ re-writes
\be
S^{EH}= \frac{\imath }{ 4 \pi G}  \int B^{AB}[e] \wedge R_{AB}[A^{CD}] + \frac{\Lambda}{3} B^{AB}[e] \wedge B_{AB}[e]\,,
\ee
corresponding to ``half'' of our action for $g=0$ and all the Lagrange multipliers vanishing.\\

Similarly, without imposing that the Plebanski 2-forms are simple but using the totally symmetric spinor-value Lagrange multiplier $\Psi_{ABCD}$, 
\be
\Psi_{ABCD}= \Psi_{(ABCD)}\,,
\ee
the generalized Plebanski action (for $g=0$) reads, 
\begin{eqnarray}
&&S^{\rm Pleb}[B^{AB},\, B^{A'B'},\, A^{MN},\, A^{M'N'},\, \Psi^{ABCD},\, \Psi^{A'B'C'D'}]= \nonumber\\
&&=\frac{\imath}{4 \pi G} \int \Big\{ B^{AB} \wedge R_{AB} [A_{MN}] - B^{A'B'} \wedge R_{A'B'} [A_{M'N'}] \nonumber\\
&&- \frac{1}{2} \Psi_{ABCD} \, B^{AB} \wedge B^{CD} + \frac{1}{2} \Psi_{A'B'C'D'} \, B^{A'B'} \wedge B^{C'D'} + \nonumber\\
&&+\frac{\lambda}{6 G}\, B^{AB} \wedge B_{AB} - \frac{\lambda}{6 G}\, B^{A'B'} \wedge B_{A'B'} \Big\} \,,
\end{eqnarray}
which is included in the action discussed in section 1.\\
 
Following these conventions, we briefly show in the next appendix some identities involved in the calculatiuons reported in the previous sections. 

\section{Identities involving the Plebanski 2-forms}

\noindent 
We can start from the very definition of self-dual and anti-self-dual variables and find 
\ba
&\Sigma^{CD}_{\rho \sigma} \Sigma_{CD\,\, \gamma \delta}  = -\frac{1}{2} \left[  \varepsilon^{A'B'} \Sigma^{CD}_{\rho \sigma}    \varepsilon_{A'B'}  \Sigma_{CD\, \, \gamma \delta}  \right] = -\frac{1}{2}  \left[  \Sigma^{+\,\, ab}_{\rho \sigma} \,\, \Sigma^+_{ab\,\, \gamma \delta}  \right] =\nonumber\\
&= -\frac{1}{2}  \left[ \frac{1}{2} \left( \Sigma^{ab}_{\rho \sigma} -\frac{\imath}{2} \varepsilon^{ab}_{\ \ \ cd}\ \Sigma^{cd}_{ \rho \sigma }\right)  \frac{1}{2} \left( \Sigma_{ab\,\, \gamma \delta} -\frac{\imath}{2} \varepsilon_{ab}^{\ \ \ rs}\ \Sigma_{rs\,\, \gamma \delta}\right)  \right] = \nonumber\\
&= \frac{1}{8}  \left[  \frac{3}{4} \Sigma^{ab}_{\rho \sigma} \, \Sigma_{ab \,\, \gamma \delta} - \imath  \Sigma^{ab}_{\rho \sigma} \varepsilon_{abcd} \,\Sigma^{cd}_{ \gamma \delta} \right]= - \frac{3}{32} g_{\gamma [ \rho} g_{\sigma ] \delta} + \frac{\imath}{8} \epsilon_{\rho \sigma \gamma \delta} \,,
\ea
having used in the first line the Infeld-van der Waerden map.\\

%\subsection{Identities involving the gravitational Hodge dual}

\noindent 
Given the 2-forms with internal indices (no matter if dual of anti-self-dual) $A^{AB}_{\mu \nu}$ and $B_{\mu\nu}^{AB}$, we find 
\ba
&&A^{AB}_{\mu \nu} B_{AB \,\,\alpha \beta} \  \epsilon^{\alpha \beta \gamma \delta} \ \epsilon^{ \mu \nu \rho \sigma} \, \left[  -\frac{3}{32} g_{\gamma [ \rho} g_{\sigma ] \delta} + \frac{\imath}{8} \epsilon_{\rho \sigma \gamma \delta} \right]= \nonumber\\ &&= 
-\frac{3}{32} A^{AB}_{\mu \nu} B_{AB \,\,\mu \nu} g^{\alpha [ \mu} g^{\nu ] \beta} + \frac{\imath}{8} A^{AB}_{\mu \nu} B_{AB \,\,\mu \nu} \epsilon^{\alpha \beta \mu \nu } = \nonumber\\
&&=-\frac{3}{32} A^{AB}_{\mu \nu} B_{AB}^{\mu \nu} + \frac{\imath}{8} A^{AB}_{\mu \nu} \star  B_{AB}^{\mu \nu} \,.
\ea
Similarly for $A^{A'B'}_{\mu \nu}$ and $B_{\mu\nu}^{A'B'}$, we find
\ba
&&A^{A'B'}_{\mu \nu} B_{A'B' \,\,\alpha \beta} \  \epsilon^{\alpha \beta \gamma \delta} \ \epsilon^{ \mu \nu \rho \sigma} \, \left[  -\frac{3}{32} g_{\gamma [ \rho} g_{\sigma ] \delta} + \frac{\imath}{8} \epsilon_{\rho \sigma \gamma \delta} \right]= \nonumber\\
&&= -\frac{3}{32} A^{A'B'}_{\mu \nu} B_{A'B' \,\,\mu \nu} g^{\alpha [ \mu} g^{\nu ] \beta} + \frac{\imath}{8} A^{A'B'}_{\mu \nu} B_{A'B' \,\,\mu \nu} \epsilon^{\alpha \beta \mu \nu } = \nonumber\\
&&=-\frac{3}{32} A^{A'B'}_{\mu \nu} B_{A'B'}^{\mu \nu} + \frac{\imath}{8} A^{A'B'}_{\mu \nu} \star  B_{A'B'}^{\mu \nu} \,.
\ea

Notice that we have used the definition of the Levi-Civita tensors $\epsilon$ written in terms of the Levi-Civita symbols $\varepsilon$, such that $\varepsilon^{0123}=\varepsilon_{0123}=1$, $\epsilon^{\alpha \beta \gamma \delta}=e^{-1}\varepsilon^{\alpha \beta \gamma \delta}$ and $\epsilon_{\alpha \beta \gamma \delta}=e\, \varepsilon_{\alpha \beta \gamma \delta}$. We have also used the definition of the gravitational Hodge dual $\star$ introduced in the previous appendix. \\

%\subsection{Identities involving the determinant of the metric}

\noindent
Below, we show some identities concerning or involving the determinant of the metric. From the very definition of the $\Sigma^{ab}$ 2-forms it is straightforward to check that
\be
\varepsilon_{abcd} \ \Sigma^{ab}_{\mu \nu} \, \Sigma^{cd}_{\rho \sigma} \, = \epsilon_{\mu\nu\rho\sigma}\,.
\ee 
For any two 2-forms $A_{ab}$ and $B_{cd}$ that can be decomposed in terms of  tensors $A_{AB}$ and $B_{CD}$ symmetric in the spinorial indices, using the Infeld-van der Waerden map and defining $( \dots ) \cdot \epsilon =  ( \dots )_{\alpha \beta \gamma \delta } \ \  \epsilon^{\alpha \beta \gamma \delta } $, we find that
\ba
& \imath \ \left(  A_{AB} \wedge B_{CD}  \right) \cdot \epsilon \ \  \Sigma^{AB} \wedge   \Sigma^{CD} =
\imath \ \left(  A_{AB} \wedge B_{CD}  \right) \cdot \frac{\varepsilon}{e} \  \left( \Sigma^{AB} \wedge   \Sigma^{CD} \right) \cdot \frac{\varepsilon}{e}  \ \ e d^4 x = \nonumber\\
&= \ \ A_{ab\,\, \mu\nu} \  B_{cd\,\, \rho\sigma} \ \ \Sigma^{ab}_{\alpha \beta} \Sigma^{cd}_{\gamma \delta}\,\, \epsilon^{\alpha \beta \gamma \delta} \ \ \epsilon^{\mu\nu \rho \sigma} \ e d^4 x= A_{ab\,\, \mu\nu} \  B_{cd\,\, \rho\sigma} \ \ \varepsilon^{abcd} \,\, \epsilon^{\mu\nu \rho \sigma} \ e d^4 x\,.
\ea
We can also prove that $B_{AB}\wedge B^{AB} = 4! \ \imath \  e \ d^4x + O(g^2)$, since 
\ba
4! \ e\  d^4x &=& 
(e_a \wedge e_b \wedge e_c \wedge e_d) \ \varepsilon^{abcd} =  (e_{AA'} \wedge e_{BB'} \wedge e_{CC'} \wedge e_{DD'})\ \varepsilon^{ABCD A'B'C'D'} = \nonumber\\
&=& (e_{AA'} \wedge e_{BB'} \wedge e_{CC'} \wedge e_{DD'}) \ \imath \ ( \varepsilon^{ AC}  \varepsilon^{BD }  \varepsilon^{A'D' }  \varepsilon^{B'C' } -  \varepsilon^{ AD}  \varepsilon^{BC }  \varepsilon^{A'C' }  \varepsilon^{B'D' } ) =\nonumber\\
&=& \imath \ \left[ - \left( e^{A}_{\ C'} \wedge e^{\ C'}_{B}  \right) \wedge \left( e_{A}^{\ D'} \wedge e_{\ D'}^{B} \right) -  \left( e_{C}^{\ A'} \wedge e^{D}_{\ A'}  \right) \wedge \left( e^{C}_{\ B'} \wedge e_{D}^{\ B'} \right)  \right] =\nonumber\\
&=& - \ \imath \ B^{AB} \wedge B_{AB} + O(g^2)\,.
\ea
Thus we find for $W$
\ba
W:& = &B_{AB} \wedge B^{AB} - B_{A'B'} \wedge B^{A'B'} = \\
& = & 4!\ \imath \ e\,  d^4 x + 2 g^2\, b_{AB} \wedge \Sigma^{AB} + g^4 \, b_{AB} \wedge b^{AB}+\nonumber\\
& \phantom{a} &- \pi^2 G^2 g^4\, \left( \delta_\xi \, 1\!\!1 + \gamma_\xi \, \star \right) F_{A'B'} \wedge \left( \delta_\xi \, 1\!\!1 + \gamma_\xi \, \star \right) F^{A'B'}  + O(g^8)  = \nonumber\\
& = & 4! \ \imath \ e \, \Bigg[ 1 -\imath \frac{g^2}{12} \, \left( {b}^{AB}  \wedge \Sigma_{AB} \right) \cdot \epsilon -\imath \frac{g^4}{24} \, \left( {b}^{AB}  \wedge b_{AB} \right) \cdot \epsilon + \nonumber \\
& \phantom{a} & + \imath \,\frac{\pi^2 G^2 g^4}{24} \, \left( \left( \delta_\xi \, 1\!\!1 + \gamma_\xi \, \star \right) F_{A'B'} \wedge \left( \delta_\xi \, 1\!\!1 + \gamma_\xi \, \star \right) F^{A'B'} \right) \cdot \epsilon  \Bigg]  \ d^4 x + O(g^8)\,,
\ea
and consequently 
\ba
\frac{1}{W}&\equiv& \frac{1}{W\cdot \epsilon} =\\
&=&  \frac{1}{4! \ e \ \imath} \ \Bigg[ 1 +\imath \frac{g^2}{12} \, \left( {b}^{AB}  \wedge \Sigma_{AB} \right) \cdot \epsilon +\imath \frac{g^4}{24} \, \left( {b}^{AB}  \wedge b_{AB} \right) \cdot \epsilon + \nonumber \\
& \phantom{a} & - \imath \,\frac{\pi^2 G^2 g^4}{24} \, \left( \left( \delta_\xi \, 1\!\!1 + \gamma_\xi \, \star \right) F_{A'B'} \wedge \left( \delta_\xi \, 1\!\!1 + \gamma_\xi \, \star \right) F^{A'B'} \right) \cdot \epsilon  \Bigg]  \ d^4 x + O(g^8)\,, \nonumber
\ea
having used the {\it ansatz} on the shifted relation between $B^{A'B'}$ and $F^{A'B'}$.

\subsection*{Acknowledgments} 
We acknowledge Roberto Percacci for very useful discussions and comments and Kirill Krasnov, Fabrizio Nesti and Simone Speziale for suggestions and 
encouragement. We are grateful to Sergei Alexandrov for pointing out an error  in the first version of this paper.
Research at Perimeter Institute is supported by the Government of 
Canada through Industry Canada and by the Province of Ontario through the Ministry of Economic Development and Innovation.  SA and AM acknowledge the generous support of the NSF CAREER grant and Dartmouth College's Dean of Faculty office.

\end{document}